\documentclass[11pt]{article}

\usepackage[letterpaper,margin=1in]{geometry}
\usepackage{amsmath,amssymb}
\usepackage{graphicx}
\graphicspath{{figures/}}   
\usepackage{textcomp}
\usepackage[super,sort&compress]{natbib}   
\usepackage{caption}

\newcommand{\ee}{\mathrm{e}}
\newcommand{\Ar}{\mathrm{Ar}}
\newcommand{\diff}[2]{\frac{d#1}{d#2}}

\begin{document}

\begin{center}
{\small This paper was a non-published draft of a research paper carried out in the mid-1980's and constitutes Appendix H in the thesis of M.A.Cappelli, 1987.}\\[2.2em]
{\large Plasma Channel Formation Through Laser Resonance Saturation}\\[1.6em]
M. A. Cappelli, S. K. Wong, R. S. Kissack and R. M. Measures\\[1.0em]
Institute for Aerospace Studies\\
University of Toronto\\
4925 Dufferin Street\\
Downsview, Ontario, Canada\\
M3H 5T6
\end{center}

\vspace{1.5em}

\section*{Abstract}

We have developed a computer code that models the three dimensional nature of a
sodium plasma created by laser resonance saturation. This was accomplished by
taking account of the temporal and spatial distortion suffered by the laser
pulse as it propagates through a nonuniform sodium atom distribution. We have
also undertaken the first measurements of the radial and axial electron density
and temperature profiles arising in such a plasma and have been able to
demonstrate that our computer code predictions are in reasonable agreement with
these results. Absorption of the laser pulse is found to be quite significant
for even a few centimeters of $10^{16}$~cm$^{-3}$ sodium atom density and that
as a result quite low temperature plasmas can be produced for modest energy
laser pulses.

\section*{Introduction}

\begin{figure}[htbp]
\centering
\includegraphics[width=0.85\textwidth]{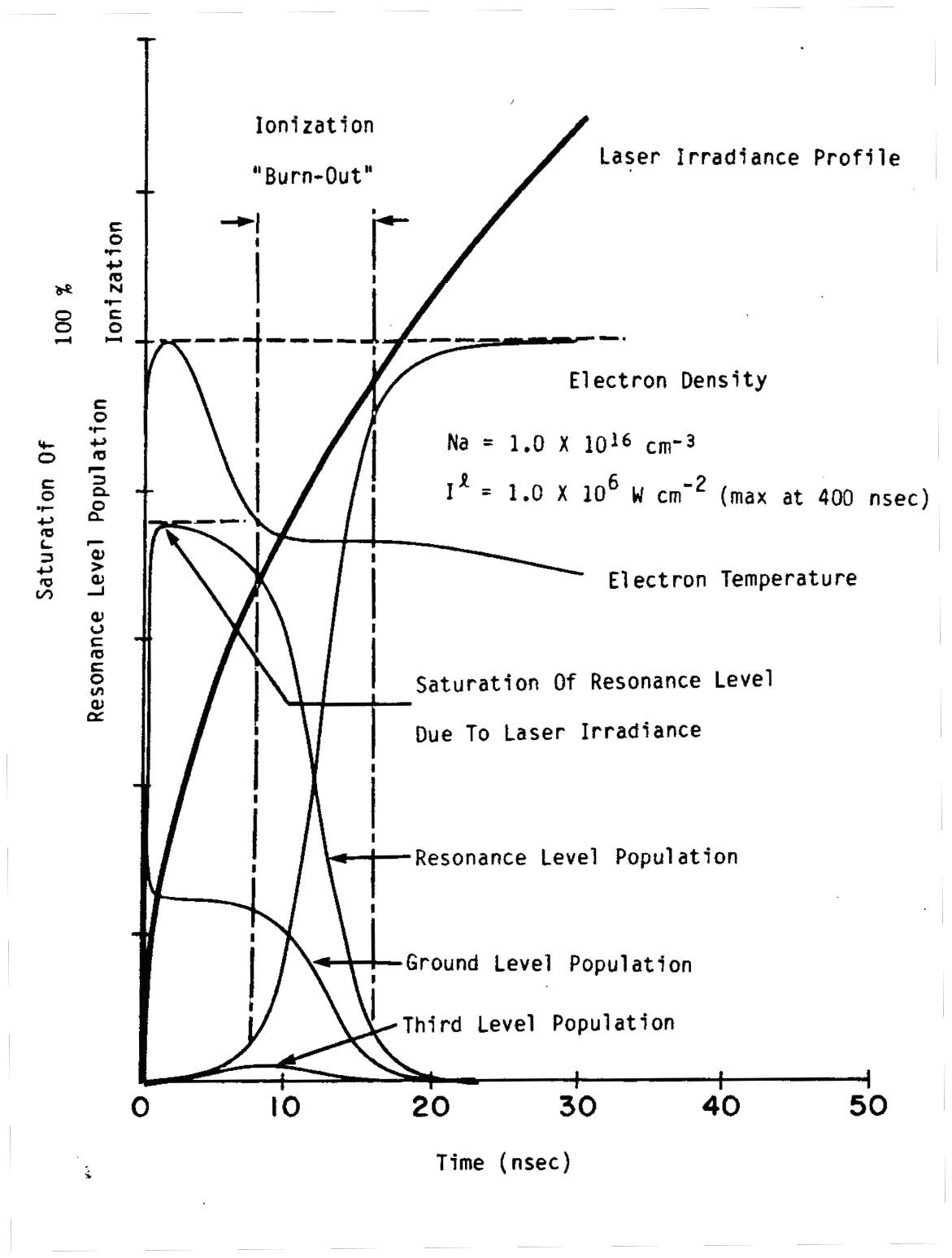}
\caption{Schematic representation of ionization and electron heating attained
through laser resonance saturation.}
\label{fig:1}
\end{figure}

Laser saturation of an atomic resonance transition clearly involves radiatively
locking the resonance and ground state populations in the ratio of their
degeneracies.\cite{ref1} Lucatorto and McIlrath\cite{ref2,ref3} were the first
to show experimentally that this interaction represents an effective method of
producing near total ionization of alkali atoms. Subsequent experiments have not
only verified this observation but have shown that this form of laser ionization
can be extended to other elements.\cite{ref4,ref5,ref6,ref7} There is growing
evidence that absorption and superelastic electron quenching collisions together
represent the primary mechanism for rapidly extracting from the laser pulse the
large amount of energy needed to achieve this high degree of
ionization.\cite{ref4,ref5,ref6,ref7,ref8,ref9,ref10,ref11,ref12,ref13,ref14,ref15,ref16}
This mechanism was first suggested by Measures,\cite{ref17, ref18} and is schematically
illustrated in figure~\ref{fig:1}. A useful overview of the subject was provided
by Jahreiss and Huber.\cite{ref5}

Measures and coworkers,\cite{ref19,ref20,ref21} have proposed that laser
ionization based on resonance saturation (LIBORS) proceeds in four stages: (i)
the laser rapidly ``locks'' the ground and resonance level populations roughly
in the ratio of their respective degeneracies, (ii) there is a growth of free
electrons due to one or more ``seed ionization'' processes and these electrons
very quickly acquire energy through ``superelastic collision quenching'' of the
laser maintained resonance state population, (iii) these ``hot'' electrons
proceed to collisionally excite and ionize the resonance state atoms and in
addition the collisionally excited higher lying levels are ``photoionized'' by
the laser field, finally (iv) ``runaway'' collisional ionization of the upper
levels occurs once some critical free electron density is achieved and this
leads to near complete ionization burnout of the species.

However, it should be noted that if a very high rate of seed ionization exists
the four stages telescope to three and the burnout stage precedes almost
directly from seed ionization.\cite{ref19}

During the ionization burnout stage superelastic heating can no longer balance
collisional cooling due to depletion of the resonance state population and
consequently the electron temperature is predicted\cite{ref19} to fall
significantly. This effect has also been predicted by Morgan\cite{ref22} who has
modelled the ionization using the Boltzmann equation rather than assuming that
the free electrons have a Maxwellian velocity distribution.

\begin{figure}[htbp]
\centering
\includegraphics[width=0.85\textwidth]{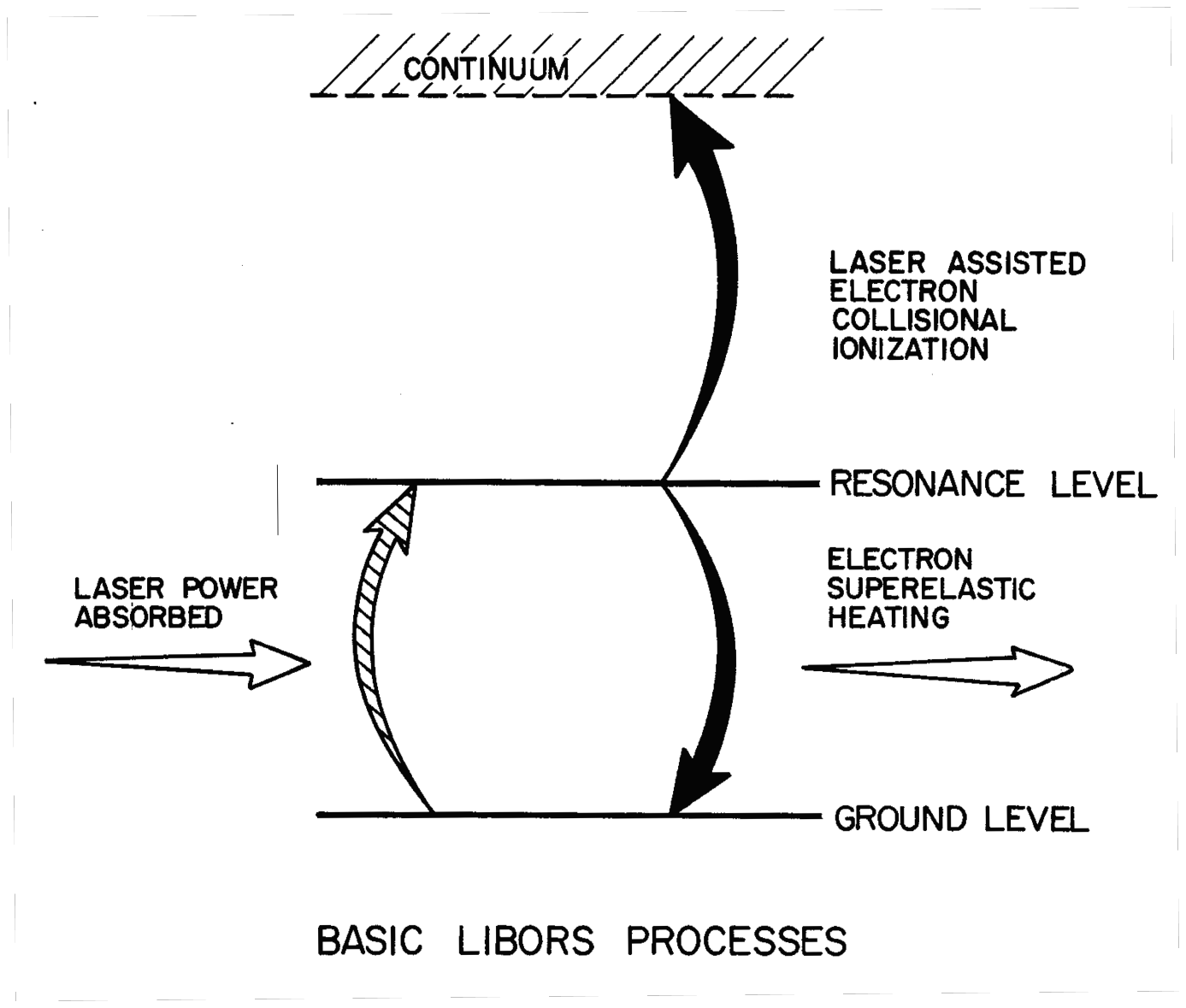}
\caption{Temporal variation of various parameters resulting from laser resonance
saturation.}
\label{fig:2}
\end{figure}

In the original formulation of our LIBORS collisional--radiative computer code
the sodium atom was represented by a 20-atomic level model and we assumed a
step-like laser pulse.\cite{ref19,ref20,ref21,ref23} An example of the temporal
behaviour of the free electron density, free electron temperature, resonance
state density, and ground state density predicted by this code modified to
include a realistic flashlamp pumped dye laser pulse (such as used in the early
experiments\cite{ref6,ref7}) is presented in figure~\ref{fig:2}. Also shown on
this figure is the assumed temporal growth of the laser irradiance. The
influence of molecular nitrogen on this ionization process has also been
considered by Measures et al\cite{ref24} and Wong.\cite{ref25} In an actual
experiment the laser pulse propagates through a non-uniform atom density
distribution and suffers appreciable absorption due to the strong nature of the
interaction. This can lead to a considerable temporal and spatial distortion of
the laser pulse. Consequently, the atoms at different locations along the path of
the laser pulse are subject to progressively less energy and to a shorter period
of excitation.

In order to better model the real effects associated with creating a plasma
channel by laser resonance saturation we have undertaken the first analysis that
takes account of both laser absorption and the non-uniform atom density
distribution along the path of the laser pulse. We have also taken into account
the spatial and temporal profile of the laser pulse.

In order to test this theoretical work we have attempted to computationally
simulate as closely as possible experiments that we have undertaken using a
specially designed sodium heat sandwich oven\cite{ref26} and a Nd-YAG laser
pumped dye laser.

\section*{5-Level Atom Model}

\begin{figure}[htbp]
\centering
\includegraphics[width=0.85\textwidth]{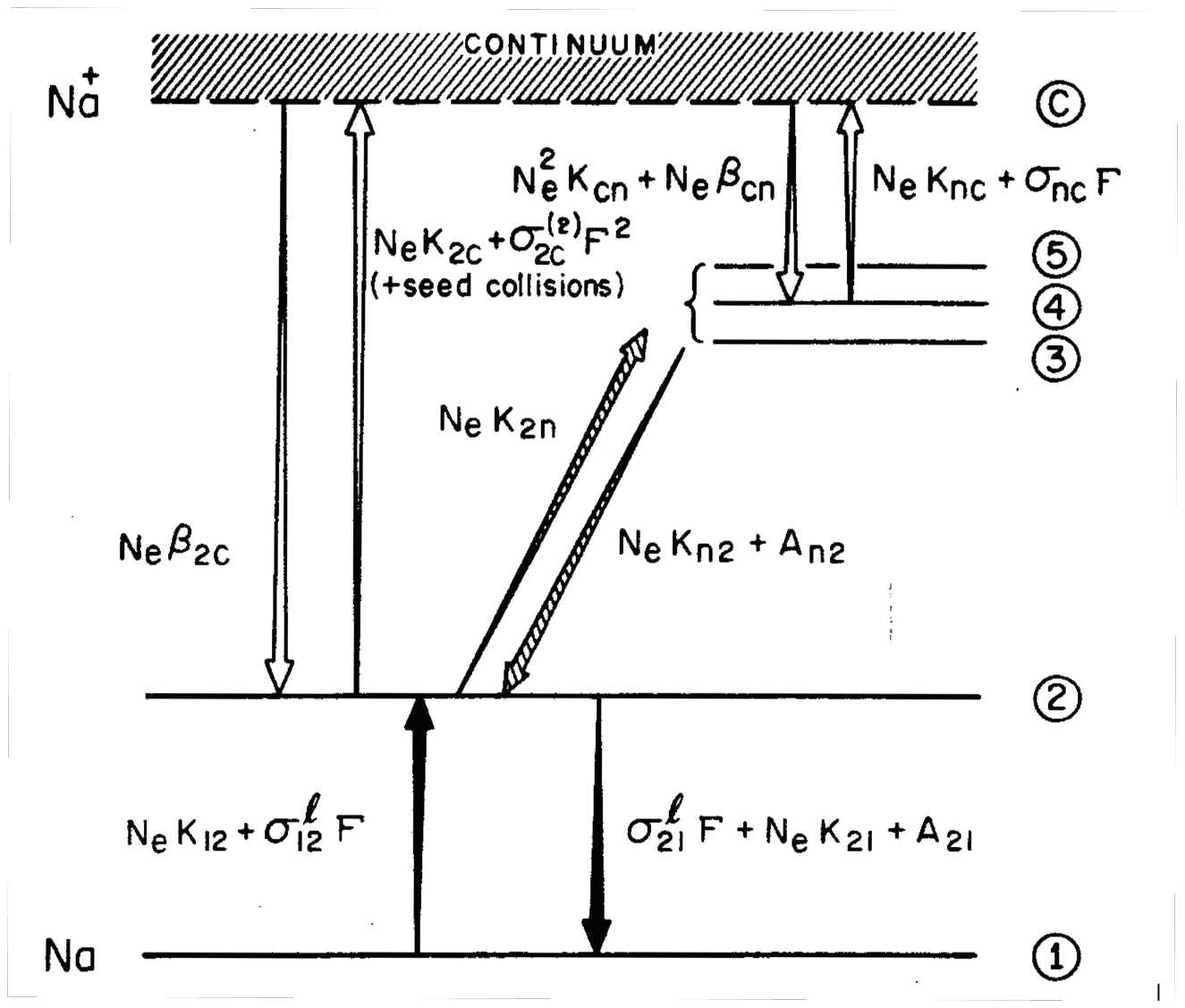}
\caption{Five level model of sodium atom.}
\label{fig:3}
\end{figure}

In an attempt to keep the analysis economically viable we replaced our 20-level
model of the sodium atom\cite{ref19,ref20,ref21,ref23} with the simplified
5-level model\cite{ref25} shown in figure~\ref{fig:3}. Although 5 levels are
indicated in figure~\ref{fig:3}, we assume that the collisional coupling between
levels 3 ($4^2S$), 4 ($3^2D$) and 5 ($4^2P$) is sufficiently rapid that their
populations are maintained in a Boltzmann equilibrium. This assumption permits us
to treat this trio of levels as a manifold and thereby reduce the number of
atomic level rate equations to three. In effect we consider only the
collisional--radiative processes ``into'' and ``out of'' this (3-4-5) level
manifold.

The remaining 15 high lying energy levels used in the 20-level model (not shown
in figure~\ref{fig:3}) are partially taken into account by assuming that
electron excitation to them (from any of the five lower levels) leads, in
effect, to immediate ionization of the atom. We therefore introduce the 5-level
``effective'' electron collisional ionization rate coefficient for each of the
low lying levels ($m = 1$ to 5) defined by the relation

\begin{equation}
K^{*}_{mc} \;=\; \sum_{n=6}^{21} K_{mn}
\label{eq:1}
\end{equation}
where $K_{mn}$ represents the electron collisional rate coefficient for the
optically allowed transition between levels $m$ and $n$. Note that $K_{m21}$ is
the $m$-level electron collisional ionization rate coefficient according to the
20-level model of the sodium atom. The corresponding ``effective'' three body
recombination rate coefficient to each of the low lying levels ($m = 1$ to 5) is
similarly constituted to take account of recombination to the higher levels,
followed by a rapid cascade down to the lower levels. We thus introduce

\begin{equation}
K^{*}_{cm} \;=\; \sum_{n=6}^{20} K_{cn}\, p(n,m)
\label{eq:2}
\end{equation}
where $K_{cn}$ represents the three-body recombination rate coefficient to level
$n$ used in our 20-level LIBORS code and $p(n,m)$ is introduced as a factor that
attempts to reflect the effective probability that a captured free electron
arriving in level $n$ subsequently decays to level $m$. In principle radiative
recombination could be included into the effective recombination rates but for
our conditions its omission was found to lead to negligible
error.\cite{ref33}

In calculating the electronic excitation rate coefficients we have used Seaton
cross-sections\cite{ref27} for the optically allowed transitions, and Gryzinski
cross-sections\cite{ref28} for all other transitions, and a Maxwellian velocity
distribution. The temperature dependence of $K_{21}$, however, was based on the
tabulated values of Crandall et al.\cite{ref29} Although both Seaton and
Gryzinski collision rates were employed for transitions between the 5 lower
levels only Seaton collision rates were used for transitions that terminated
above level 5.

We have found that close agreement can be obtained between the 5 and 20-level
LIBORS codes if these probability factors are generally given by the expression

\begin{equation}
p(n,m) \;=\; \frac{N_e K_{nm} + A_{nm}}
{N_e K_{nc} + \displaystyle\sum_{n<5}\bigl(N_e K_{nq} + A_{nq}\bigr)}
\label{eq:3}
\end{equation}
where $m = 1$ to 5, $N_e$ is the free electron density, $A_{nm}$ and $A_{nq}$ are
the Einstein spontaneous transition probabilities between levels $n$ and $m$ and
$n$ and $q$ respectively, with $q$ being the levels that are optically connected
to level $n$. To simplify the computation, only those $K_{nc}$ terms from levels
within 0.5~eV of the continuum are included in the calculation of $p(n,m)$.

The 5-level model can thus be described by the following set of equations: first
we have the three atomic level population rate equations,

\begin{equation}
\begin{split}
\diff{N_1}{t} \;=\;& \bigl(N_2 - g N_1\bigr) R_{21}
 + N_2 C_{21} + N_5 C_{51}
 + N_e\bigl(N_3 K_{31} + N_4 K_{41}\bigr) + N_e^{3} K^{*}_{c1} \\[2pt]
 & - N_1\left[\,N_e\left(\sum_{m=2}^{5} K_{1m} + K^{*}_{1c}\right)\right]
\end{split}
\label{eq:4}
\end{equation}

\begin{equation}
\begin{split}
\diff{N_2}{t} \;=\;& \bigl(g N_1 - N_2\bigr) R_{21}
 + N_e\bigl(N_1 K_{12} + N_5 K_{52}\bigr)
 + N_3 C_{32} + N_4 C_{42} + N_e^{3} K^{*}_{c2} \\[2pt]
 & - N_2\left[\,N_e\left(\sum_{m=3}^{5} K_{2m} + K^{*}_{2c}\right) + C_{21}\right]
   - N_2^{2} v \bigl(\sigma_L F + \sigma_A\bigr)
   - N_2 \sigma^{(2)}_{2c} F^{2}
\end{split}
\label{eq:5}
\end{equation}

and

\begin{equation}
\begin{split}
\diff{N^{*}_{3}}{t} \;=\;& N_e\left[\sum_{m=3}^{5}
 \bigl(N_1 K_{1m} + N_2 K_{2m} + N_e^{2} K^{*}_{cm}\bigr)\right] \\[2pt]
 & - N_3\bigl(C_{32} + N_e K_{31} + P_{3c}\bigr)
   - N_4\bigl(C_{42} + N_e K_{41} + P_{4c}\bigr)
   - N_5\bigl(C_{51} + N_e K_{52} + P_{5c}\bigr)
\end{split}
\label{eq:6}
\end{equation}
where we have introduced the (3-4-5) level manifold population density

\begin{equation}
N^{*}_{3} \;=\; N_3 + N_4 + N_5
\label{eq:7}
\end{equation}

As indicated we assume that the close spacing of these three levels ensures
collisional equilibrium and consequently that the internal distribution of this
manifold is described by a Boltzmann distribution. In which case

\begin{equation}
N_j \;=\; N^{*}_{3} \big/ U_j(T_e)
\label{eq:8}
\end{equation}
where

\begin{equation}
U_j(T_e) \;=\; \sum_{m=3,4,5} g_{mj}\, \ee^{-E_{mj}/kT_e}
\label{eq:9}
\end{equation}
with $g_{mj} = g_m/g_j$ and $E_{mj} = E_m - E_j$ ($j = 3$, 4 or 5). In order to
streamline the equations as far as possible we also introduced the notation,

\begin{equation}
C_{nm} \;=\; N_e K_{nm} + A_{nm}
\label{eq:10}
\end{equation}
and

\begin{equation}
P_{nc} \;=\; \sigma_{nc} F + N_e K^{*}_{nc}
\label{eq:11}
\end{equation}
where $\sigma_{nc}$ is the single photon ionization cross section for level $n$
and $F$ is the laser photon flux density, defined by

\begin{equation}
I^{\ell}(\nu - \nu_{\ell},\, t) \;=\; h\nu_{\ell}\, F(t)\, L^{\ell}(\nu - \nu_{\ell})
\label{eq:12}
\end{equation}
where $I^{\ell}(\nu-\nu_{\ell}, t)$ is the laser spectral irradiance symmetric
about line centre frequency $\nu_{\ell}$ at time $t$, $h\nu_{\ell}$ the laser
photon energy, and $L^{\ell}(\nu - \nu_{\ell})$ the laser spectral distribution.
In the above equations we have also introduced the resonance to ground level
degeneracy ratio $g$ and the stimulated emission rate per resonance state atom,

\begin{equation}
R_{21}(\Delta\nu_{\ell},\, t) \;=\; \frac{B_{21}}{4\pi}
\int_{-\infty}^{\infty} I^{\ell}\bigl(\Delta\nu_{\ell} - \Delta\nu,\, t\bigr)
L_{21}(\Delta\nu)\, d\Delta\nu
\label{eq:13}
\end{equation}
where $B_{21}$ is the Milne stimulated emission coefficient and
$L_{21}(\Delta\nu)$ is the atomic resonance line profile function,
$\Delta\nu = \nu - \nu_0$, $\nu_0$ being the line centre frequency of the
resonance transition and $\Delta\nu_{\ell}$ the laser detuning from the atomic
line centre frequency (we shall further discuss this later).

The appropriate electron energy equation can be expressed in the following form:

\begin{equation}
\begin{split}
\diff{}{t}\bigl(N_e E_e\bigr) \;=\;&
 N_e \sum_{n=2}^{5}\ \sum_{\substack{m=1\\(n \neq m)}}^{2} N_n K_{nm} E_{nm}
 \;+\; N_e^{3} \sum_{m=1}^{5} K^{*}_{cm} E'_{cm}
 \;+\; \sum_{m=3}^{5}\bigl(E_{21} - E_{cm}\bigr) N_m \sigma_{mc} F \\[2pt]
 &- N_e \sum_{m=1}^{2}\ \sum_{\substack{n=2\\(n \neq m)}}^{5} N_m K_{mn} E_{nm}
  \;-\; N_e \sum_{m=1}^{5} N_m K^{*}_{mc} E_{cm}
  \;-\; N_e^{2} H_{ei} \;-\; N_e N_{\Ar} H_{e\Ar} \\[2pt]
 &- N_e\bigl[N_1 H_{e1} + N_2 H_{e2}\bigr] \;+\; S
\end{split}
\label{eq:14}
\end{equation}
where $E_e$ represents the free electron mean thermal energy (taken as
$\tfrac{3}{2}kT_e$), and $H_{ei}$, $H_{e1}$, $H_{e2}$ and $H_{e\Ar}$ represent
the rates of elastic energy transfer to ions through Coulombic collisions, and to
Na(3s), Na(3p), and argon atoms through hard sphere collisions.\cite{ref31} $S$
corresponds to the source of electron energy arising from the various seed
ionization processes, i.e.,

\begin{equation}
S \;=\; \bigl(2E_{21} - E_{c2}\bigr)
\left[N_2 \sigma^{(2)}_{2c} F^{2} + \tfrac{1}{2} N_2^{2} v\, \sigma_L F\right]
\;+\; \tfrac{1}{2} N_2^{2} v\, \sigma_A E_A
\label{eq:15}
\end{equation}
Here $\sigma^{(2)}_{2c}$ represents the resonance state two photon ionization
coefficient, $\sigma_L$ the laser assisted Penning ionization coefficient, and
$\sigma_A$ the effective associative ionization cross section which is taken to
be $1.5 \times 10^{-15}$~cm$^{2}$. This represents a value that gave us a
reasonable agreement between our computational results and our experiments. It
should be noted that this value lies between the results suggested by Roussel et
al\cite{ref11} and the later work of Huennekens and Gallagher.\cite{ref11} $E_A$
is the effective energy of creation for the electrons liberated by this latter
process and $v$ represents the mean atom collisional velocity.
Equation~(\ref{eq:14}) is solved in conjunction with the ionization
($\dot{N}_e$) equation. A detailed discussion of the 20-level atomic code and a
comprehensive review of the various seed ionization processes with the relevant
cross sections used is provided by Cardinal.\cite{ref31}

It should be noted that another fine adjustment in the form of a small reduction
(of about 0.055~eV) in the three-body recombination energy was found to bring the
results of the 5-level LIBORS code into excellent agreement with the 20-level
code over a wide range of densities.\cite{ref25} This modified energy is
designated by $E'_{cm}$ in equation~(\ref{eq:14}).

The electron--ion energy exchange rate $H_{ei}$ is proportional to
$(T_e - T_i)$\cite{ref30} and therefore it is necessary to evaluate the ion
temperature. This is accomplished through simultaneously solving the ion energy
equation,

\begin{equation}
\diff{}{t}\bigl(N_i \varepsilon_i\bigr) \;=\;
 N_e^{2} H_{ei}
 - \bigl(N_0 - N_e\bigr) N_e H_{ia}
 - N_{\Ar} N_e H_{i\Ar}
 + \varepsilon_a \dot{N}_i
 - \varepsilon_i \dot{N}_i
\label{eq:16}
\end{equation}
where we assume that $N_e = N_i$ and $\varepsilon_i$ and $\varepsilon_a$ are the
mean thermal energies of the ion and atom, respectively. The second and third
terms in equation~(\ref{eq:16}) takes account of elastic cooling collisions
between the sodium ions and the background of sodium and argon buffer gas atoms.
$(N_0 - N_e)$ and $N_{\Ar}$ represent the sodium and argon atom densities ($N_0$
being the initial sodium density), while $H_{ia}$ and $H_{i\Ar}$ represent
respective rate coefficients for elastic energy exchange between the ions and the
background of neutral sodium and argon atoms\cite{ref32,ref33}

\begin{equation}
H_{i\Ar} \;=\; \sigma_{i\Ar}\,
 \frac{4 m_i m_{\Ar}}{\bigl(m_i + m_{\Ar}\bigr)^{2}}
 \left[\frac{8kT_i}{\pi m_i} + \frac{8kT_{\Ar}}{\pi m_{\Ar}}\right]^{1/2}
 k\bigl(T_i - T_{\Ar}\bigr)
\label{eq:17}
\end{equation}
where we have assumed the elastic energy transfer cross-section,
$\sigma_{i\Ar}$, is constant with respect to the relative sodium ion--argon
velocity, and the velocity distributions for the colliding species are
Maxwellian. $T_{\Ar}$ is the argon atoms' translational temperature. A similar
expression is used in the case of $H_{ia}$, with $m_i = m_a$ in this instance.
The last two terms in Eq.~(\ref{eq:16}) takes account of the net gain or loss of
energy to the ions resulting from their creation from atoms or their conversion
back to atoms.

This background density of argon atoms varies across the sodium vapor region such
that the sum of the partial pressures equals the pressure of the argon in the
water cooled peripheral region. The initial vapor temperature at each location is
deduced from the optically measured sodium density profile using the vapor
pressure curves provided by Nesmeyanov.\cite{ref34} By a similar argument to that
given above it follows that since $H_{iA}$ is proportional to the difference in
temperature between the sodium ions and the argon atoms, the argon temperature
needs in principle to be computed. However, we have assumed that the argon's
temperature is not changed appreciably. It should also be noted that an
electron--argon elastic energy loss term was introduced into the free electron
energy equation. To complete the set of energy equations we have also included an
equation for the mean thermal energy of the neutral sodium atoms.

Combining equations~(\ref{eq:12}) and (\ref{eq:13}) enables us to write the
stimulated rate of emission per atom in level 2 in the form

\begin{equation}
R_{21}(t) \;=\; \frac{A_{21}\lambda^{2}}{8\pi}\, F(t)
\int_{-\infty}^{\infty} L^{\ell}\bigl(\Delta\nu_{\ell} - \Delta\nu\bigr)
L_{21}(\Delta\nu)\, d\Delta\nu
\label{eq:18}
\end{equation}
where $\lambda$ is the appropriate wavelength.

In reality level 2 for the sodium atom comprises the two fine structure levels:
$3^2P_{3/2}$ and $3^2P_{1/2}$. The corresponding doublet line profile function
for the sodium resonance lines,

\begin{equation}
L_{21} \;=\; \tfrac{2}{3} L_{3/2}\bigl(\Delta\nu_{3/2}\bigr)
 + \tfrac{1}{3} L_{1/2}\bigl(\Delta\nu_{1/2}\bigr)
\label{eq:19}
\end{equation}

We now define $\Delta\nu_j = \nu - \nu_j$ ($j = 1/2$, $3/2$) where $\nu_j$
represents the line center frequency for the $(3^2P_j - 3^2S_{1/2})$ transition.
$L_{3/2}$ and $L_{1/2}$ are the respective line profiles of the individual
resonance transitions. We have assumed a Lorentzian line profile function for
each transition of the form

\begin{equation}
L_j\bigl(\Delta\nu_j\bigr) \;=\; \frac{1}{\pi}\,
 \frac{\gamma_j}{\bigl(\Delta\nu_j\bigr)^{2} + \gamma_j^{2}}
\label{eq:20}
\end{equation}
where $\gamma_j$ is the appropriate HWHM which includes contributions from both
Stark and resonance broadening.\cite{ref25,ref26,ref27, ref35}

The laser spectral distribution is approximated by a Gaussian, i.e.,

\begin{equation}
L^{\ell}\bigl(\nu - \nu_{\ell}\bigr) \;=\;
 \frac{1}{\beta_{\ell}\sqrt{\pi}}\,
 \ee^{-\left[\left(\nu - \nu_{\ell}\right)/\beta_{\ell}\right]^{2}}
\label{eq:21}
\end{equation}
where $\beta_{\ell}$ represents the Gaussian half width. The convolution of the
two profiles in (\ref{eq:18}) can be expressed in the form of a sum of two Voigt
functions which are expanded in a Taylor series to first order in
$\gamma_j/\beta_{\ell}$.\cite{ref33} The form of equation~(\ref{eq:18}) leads us
to introduce an ``effective cross-section'' for stimulated emission of the
resonance doublet,

\begin{equation}
\sigma^{\ell}_{21} \;=\; \frac{A_{21}\lambda^{2}}{8\pi}
\sum_{j=1/2}^{3/2} \int_{-\infty}^{\infty}
 L^{\ell}\bigl(\Delta\nu_{\ell} - \Delta\nu_j\bigr)
 L_j\bigl(\Delta\nu_j\bigr)\, d\Delta\nu_j
\label{eq:22}
\end{equation}

\begin{figure}[htbp]
\centering
\includegraphics[width=0.85\textwidth]{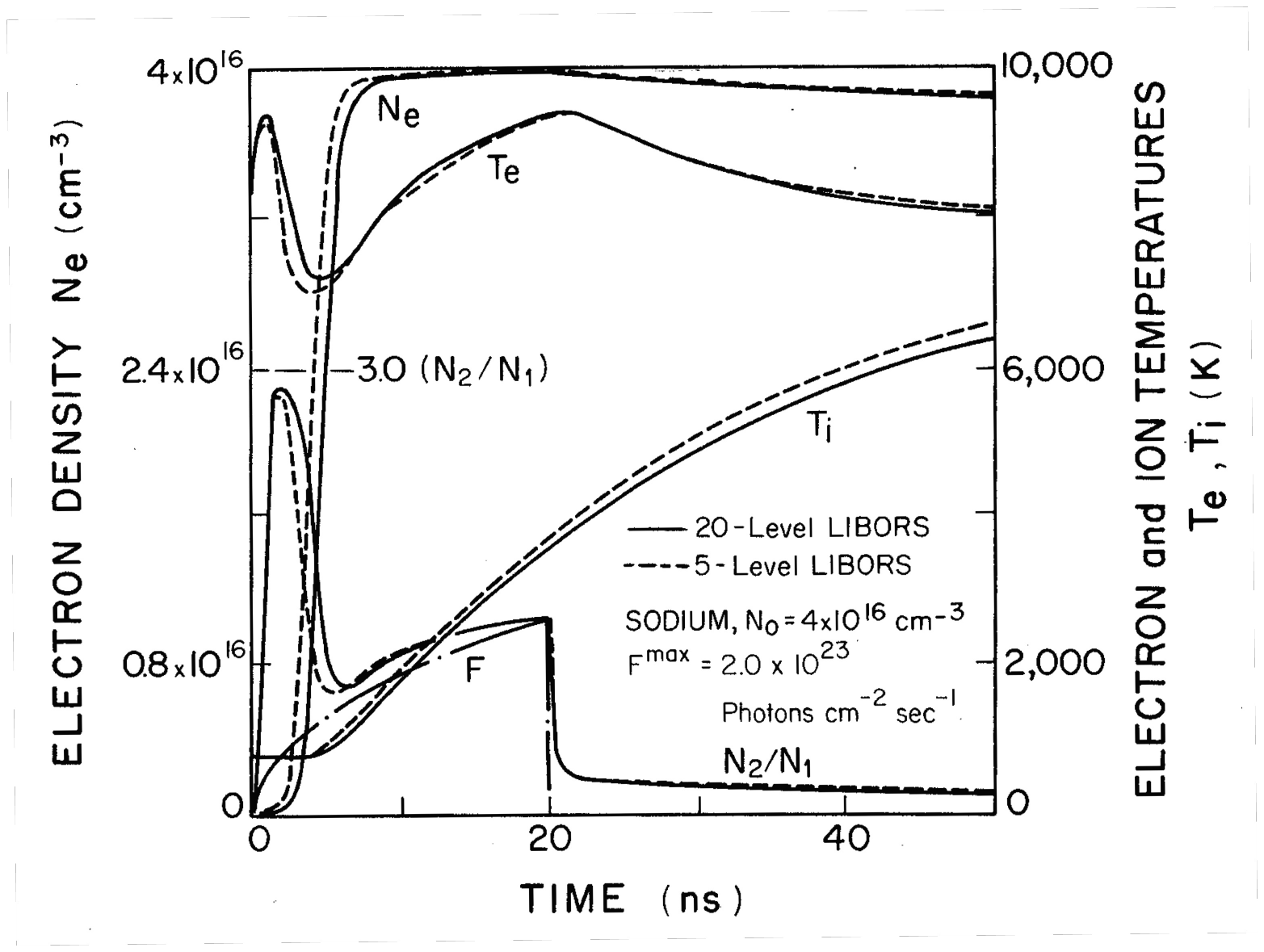}
\caption{Comparison of computer predicted temporal histories of various
parameters based on 5 and 20-level models of sodium atom
(\rule[0.5ex]{1.2em}{0.6pt}~20-level, \mbox{-\,-\,-\,-}~5-level)
$N_0 = 4\times10^{16}$~cm$^{-3}$ and
$F_p = 2\times10^{23}$~photons~cm$^{-2}$~s$^{-1}$.}
\label{fig:4}
\end{figure}

We recognized from the outset that attenuation of the laser pulse could lead to
relatively sudden changes in the laser irradiance and that in many experiments
short laser pulses are used. Consequently in our comparison of the 5 and
20-level computer codes we used the truncated laser pulse indicated as $F$ in
figure~\ref{fig:4}. With a laser pulse of this shape the comparison was tested
fairly severely. Figure~\ref{fig:4} indicates the excellent agreement obtained
between the 5 and 20-level models for a sodium density ($N_0$) of
$4 \times 10^{16}$~cm$^{-3}$. It is evident from figure~\ref{fig:4} that
$N_2/N_1$ falls from the saturated value of 3.0 at about the time of ionization
burnout even though $F$ is still increasing. This temporal decrease is attributed
to the sudden increase in the superelastic quenching rate which for the given
laser irradiance can overwhelm the stimulated emission rate as full ionization is
achieved.

It is apparent from figure~\ref{fig:4} that the 5-level LIBORS code mirrors very
closely the predictions of the 20-level code in regard to the temporal behaviour
of the parameters: $N_e$, $T_e$, $T_i$, $N_2$ and $N_1$. We feel that the small
($\lesssim 20\%$) difference in the lead time between the two sets of curves is
quite acceptable. In order to check that this close agreement was not fortuitous
to this particular density we ran comparisons at two additional densities
($N_0 = 5\times10^{15}$ and $10^{16}$~cm$^{-3}$) and found the same close
agreement.\cite{ref25}

\section*{Radiative Transfer Equation for an Extended Vapor Region}

The large cross section involved in laser resonance saturation leads to
appreciable absorption of the laser pulse as it propagates through the atomic
vapor. Consequently, any realistic simulation of plasma channel formation based
on this interaction must take account of this fact and we are therefore led to
consider the radiative transfer equation. For a laser pulse propagating in the
$z$-direction we can write

\begin{equation}
\frac{1}{c}\,\frac{\partial F}{\partial t} + \frac{\partial F}{\partial z}
 \;=\; \bigl(N_2 - g N_1\bigr) R_{21}
 - N_2 \sigma^{(2)}_{2c} F^{2}
 - \tfrac{1}{2} N_2^{2} v\, \sigma_L F
 - \sum_{m=3}^{5} N_m \sigma_{mc} F
\label{eq:23}
\end{equation}
The first term on the right hand side takes account of the net loss of laser
photons due to absorption in the resonance transition, the second term allows for
two photon ionization of the resonance level, the third term involves laser
assisted Penning ionization of the resonance level and the last term refers to
single photon ionization of the levels within the (3-4-5) level manifold.

\begin{figure}[htbp]
\centering
\includegraphics[width=0.85\textwidth]{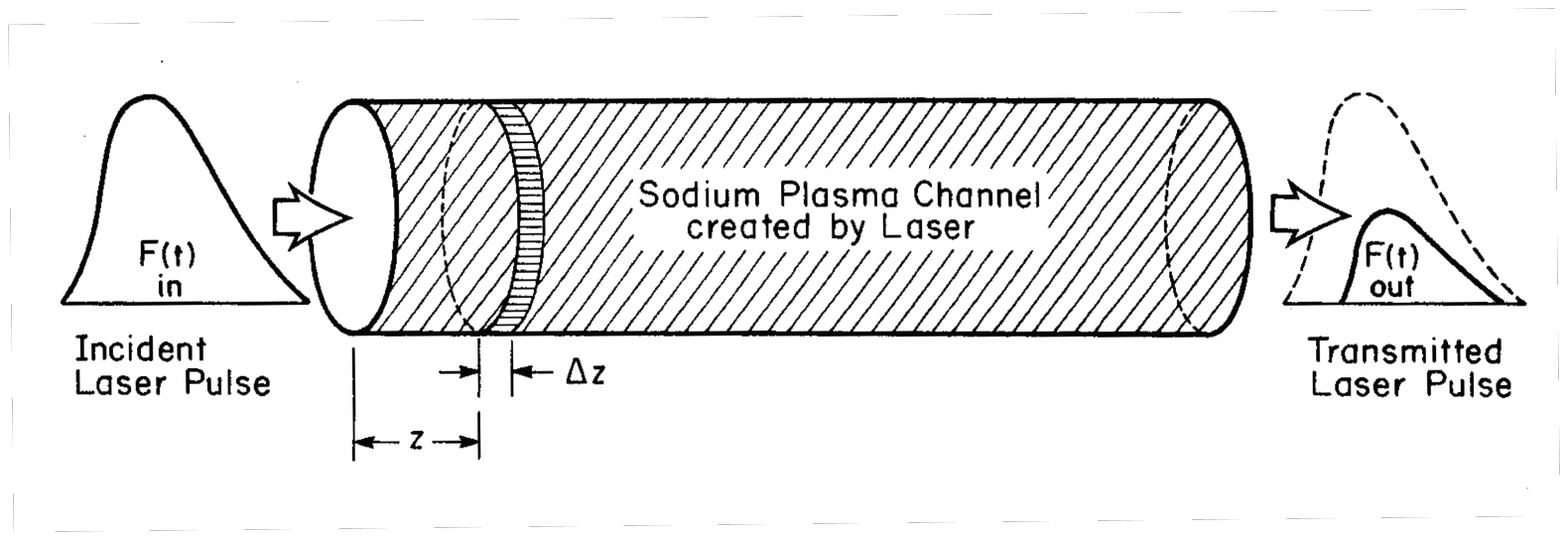}
\caption{Multislab model used for LIBORS computer calculations showing
attenuation of laser pulse.}
\label{fig:5}
\end{figure}

It is quite clear that addressing the problem of laser resonance saturation for
an extended vapor region involves the solution of the population and energy
equations indicated earlier, together with this radiative transfer equation in
both space and time. We decided to tackle this problem by dividing the irradiated
vapor column into a series of slabs, see figure~\ref{fig:5}, each of which was
assumed to have a uniform atom density equal to the value of the empirical
function describing the experimentally measured sodium atom density at the front
of the pertinent segment of vapor. The thickness of each slab was determined by
the requirements that at least 99\% of the energy fluence incident on any given
slab be transmitted to the next slab\cite{ref33} and that the maximum value be
limited to 1~mm. In reality, there would be some degree of spectral hole burning
expected as the laser pulse propagates through the vapor. However, as shown by
Measures and Herchen\cite{ref36} and Wong,\cite{ref25} the omission of this
feature may not lead to appreciable error. In view of this and in order to
minimize computing time, we have assumed that the spectral laser distribution
remains unchanged as it propagates through the vapor. Therefore, we can
generalize equation~(\ref{eq:12}), for the spectral irradiance at position $z$ to
read

\begin{equation}
I^{\ell}\bigl(\nu - \nu_{\ell},\, z,\, t\bigr) \;=\;
 h\nu_{\ell}\, F(z,t)\, L^{\ell}\bigl(\nu - \nu_{\ell}\bigr)
\label{eq:24}
\end{equation}
Moreover, for the temporal and spatial scales of interest in our experiment, the
speed of light $c$ can be treated as infinite. This fact, in conjunction with
equation~(\ref{eq:24}) implies equation~(\ref{eq:23}) can be rewritten as

\begin{equation}
\frac{\partial F}{\partial z} \;=\;
 \bigl(N_2 - g N_1\bigr) \sigma^{\ell}_{21} F
 - N_2 \sigma^{(2)}_{2c} F^{2}
 - \tfrac{1}{2} N_2^{2} v\, \sigma_L F
 - \sum_{m=3}^{5} N_m \sigma_{mc} F
\label{eq:25}
\end{equation}
where $\sigma^{\ell}_{21}$ was given by equation~(\ref{eq:22}).

Within each slab the population and energy equations are numerically integrated
(using an IMSL subroutine DGEAR\cite{ref33}), subject to the laser pulse exiting
the previous slab to provide the temporal histories of such parameters as:
$N_e$, $T_e$, $T_i$, $T_a$, $N^{*}_{3}$, $N_2$ and $N_1$. In this multislab
model, equation~(\ref{eq:24}) is solved and the laser photon flux density exiting
the $i^{\mathrm{th}}$ slab (bounded by $z_i$ and $z_{i-1}$)

\begin{equation}
F\bigl(z_{i+1},\, t\bigr) \;=\; F\bigl(z_i,\, t\bigr)\,
 \ee^{-p_i \Delta z_i}
\label{eq:26}
\end{equation}
and is seen to be expressed in terms of the incident value
$F(z_{i-1}, t)$. In this equation

\begin{equation}
p_i \;=\; \left[\bigl(g N_1 - N_2\bigr) \sigma^{\ell}_{21}
 + N_2 \sigma^{(2)}_{2c} F\bigl(z_{i-1},\, t\bigr)
 + \tfrac{1}{2} N_2^{2} v\, \sigma_L
 + \sum_{m=3}^{5} N_m \sigma_{mc}\right]_i
\label{eq:27}
\end{equation}
and

\begin{equation}
\Delta z_i \;=\; z_{i+1} - z_i
\label{eq:28}
\end{equation}

This is permissible because the loss of laser energy through two photon
ionization is small compared to that extracted through resonance pumping. The
value of $F(z_i, t)$ for each time step is stored and then used for calculating
the temporal variation of the parameters in the subsequent
$(i+1)^{\mathrm{th}}$ slab.

The incident temporal laser profile used in these calculations was chosen to
approximate the experimental laser pulse by an analytical curve of the form

\begin{equation}
F^{\ell}(0,\, t) \;=\; \varepsilon_{\ell}\, a\, t^{b}\, \ee^{-ct^{2}}
\label{eq:29}
\end{equation}
where $b$ and $c$ were constants determined from the experiment,
$\varepsilon_{\ell}$ is the energy fluence in the laser pulse and ``$a$'' is a
normalization constant determined from the relation

\begin{equation}
\int_{0}^{\tau_{\ell}} a\, t^{b}\, \ee^{-ct^{2}}\, dt \;=\; 1
\label{eq:30}
\end{equation}

\section*{Experimental Facility}

\begin{figure}[htbp]
\centering
\includegraphics[width=0.85\textwidth]{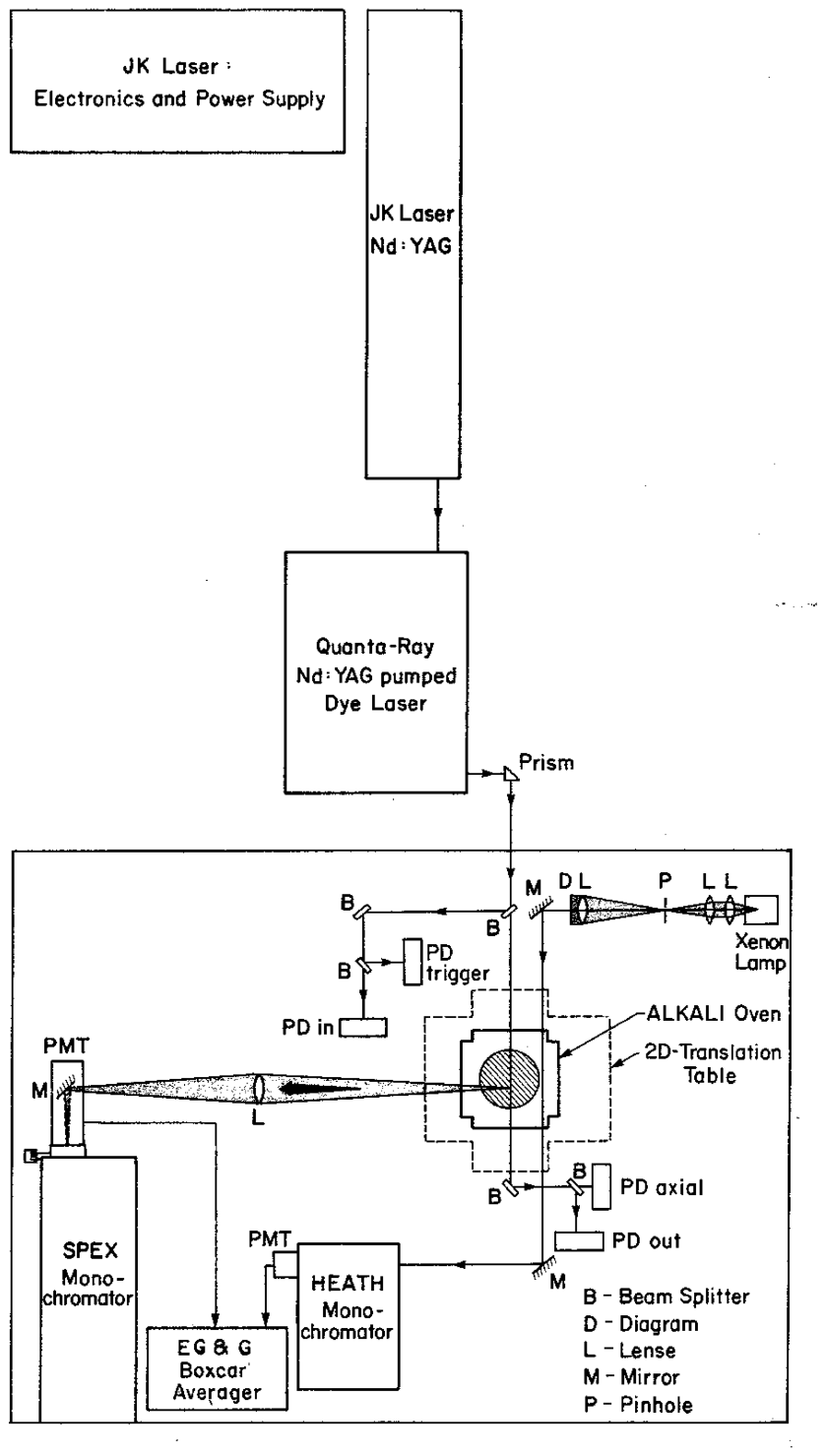}
\caption{Schematic of LIBORS experimental facility.}
\label{fig:6}
\end{figure}

The experimental facility developed for this research is schematically
illustrated in figure~\ref{fig:6}. The second harmonic of a Neodymium-YAG laser
(JK HY750) was used to pump a dye laser (Quanta-Ray PDL). Using Kiton Red-620 dye
(Exciton Chemical Co.) an output of about 40~mJ at a 10~Hz repetition rate was
obtained over a wavelength range that included the two sodium D lines at 589.0
and 589.6~nm. The dye laser's spectral full width half maximum FWHM was measured
to be about 0.02~nm.

A specially designed heat sandwich oven was used to generate a disc-shaped zone
of sodium vapor. This oven is in the form of a rectangular box with circular
heater plates and steel wicks, top and bottom, surrounded by cooling water coils
to prevent the sodium vapor from reaching the four windows that provide
360\textdegree{} optical access. A more detailed description of the method of
measuring the sodium atom density distribution is provided
elsewhere.\cite{ref26}

The electron density measurements were undertaken using a sideways mounted SPEX
1700 II monochromator. This enables the entrance slit of the monochromator to
sample a thin horizontal slab of the plasma emission at a height $y$ (scanned by
adjusting the input optics) above the axis of the laser beam. The magnification
of the optics was 1:1 with the SPEX entrance slit at 1~cm $\times$ 40~$\mu$m.

The signal from the RCA C31034 photomultiplier mounted on the SPEX monochromator
was processed by an EG\&G 4420 Signal Averager that was typically gated to sample
the emission in a 2~ns interval, some 65~ns after the start of the laser pulse.
Since the laser pulse duration was less than 40~ns, the influence of the laser
field (through the dynamic Stark effect) should be minimal and spontaneous and
electron collision induced decay of the resonance state population should ensure
that optical depth effects are small.

In order to evaluate the radial profile of the free electron density $N_e$,
spectral scans of the $4^2D$--$3^2P$ multiplet were taken for 15 lateral ($y$)
positions across the plasma column. The spectral resolution was estimated to be
0.05~nm (FWHM) with 16 laser shots being averaged to constitute one intensity
measurement (at a given wavelength) and 128 such measurements comprise one
spectral scan.

\section*{Comparison of Computational and Experimental Results}

The present form of our LIBORS computer code predicts the state of the plasma
formed along the direction of propagation for a given sodium atom distribution
and a temporally prescribed laser field. In an experiment the laser field also
has a radial variation and this has to be taken into account if we wish to model
the three dimensional nature of the interaction. To accomplish this we have
assumed that the incident laser pulse has a Gaussian radial distribution (the
justification for this is presented elsewhere\cite{ref33}), and the same temporal
history given by equation~(\ref{eq:29}), at each radial position.

\begin{figure}[htbp]
\centering
\includegraphics[width=0.85\textwidth]{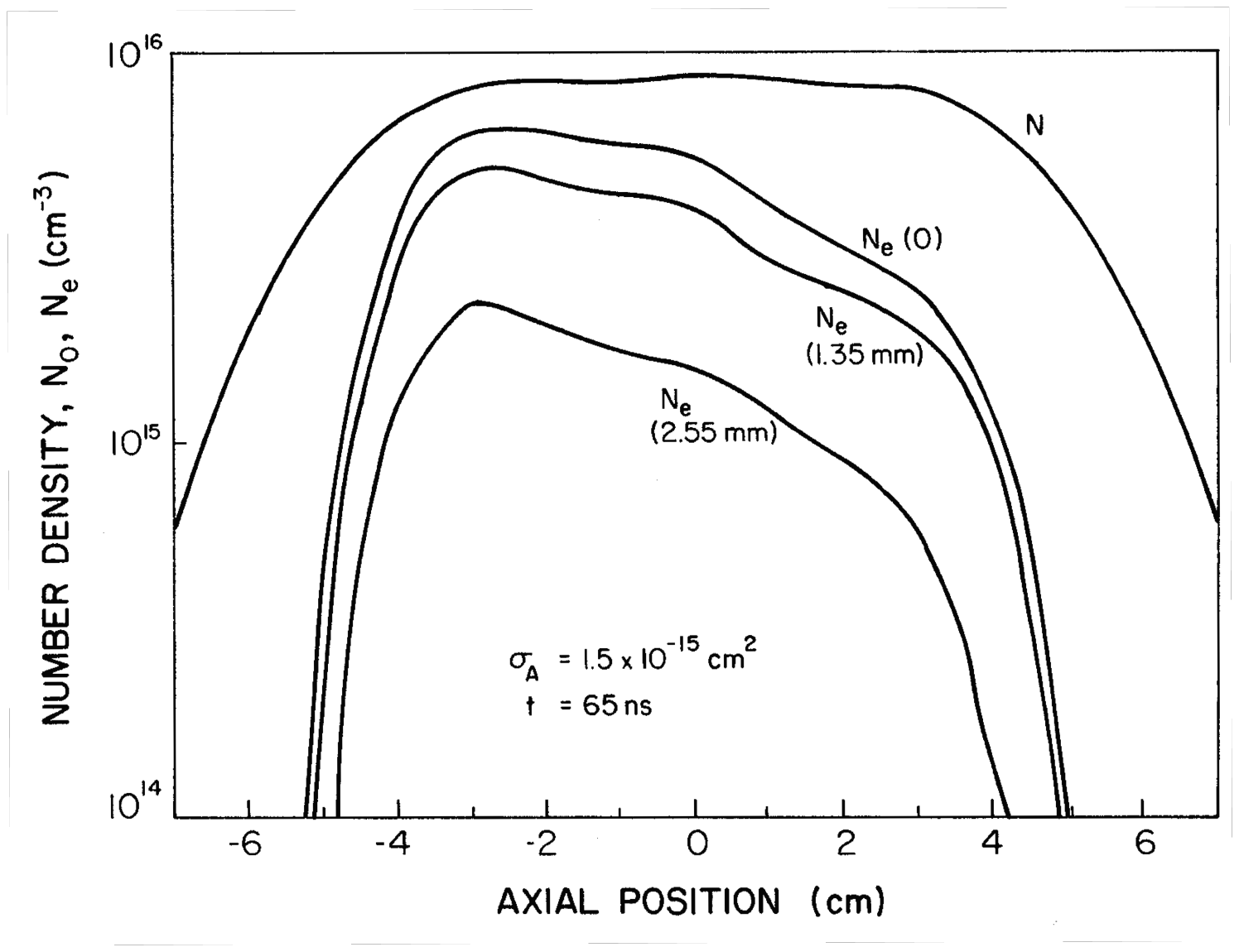}
\caption{Axial variation of free electron density $N_e(z)$ at four radial positions
($r = 0$, 1.35, and 2.55~mm) predicted by LIBORS code at $t = 65$~ns. Also
shown is the axial sodium atom density distribution $N(z)$ used in the
computation.}
\label{fig:7}
\end{figure}

A series of computer runs for a range of incident laser energy fluence was then
undertaken. The state of the plasma at any position can then be predicted from
these computer runs by assigning the appropriate laser energy fluence to each
radial position. By way of example we have displayed in figure~\ref{fig:7} the
predicted axial variation in the free electron density $N_e(z)$ corresponding to
four radial positions ($r = 0$, 1.25 and 2.55~mm) 65~ns after the start of the
laser pulse for the experimentally based sodium atom distribution $N_0(z)$ also
shown. The incident laser pulse used to generate these results was assumed to
have the temporal variation given by equation~(\ref{eq:29}) with $b = 2.64$,
$c = 6.5\times10^{-3}$ and $\tau_{\ell} = 40$~ns and a Gaussian radial profile
with an energy of 25~mJ and $r_0 = 2.5$~mm. These three radial positions
corresponded to incident laser energy fluences of: just 127, 95 and 45
(mJ~cm$^{-2}$) for this particular laser pulse.

A comparison of the free electron density predicted at equidensity axial
locations in figure~\ref{fig:7} reveals that both the electron density and the
radius of the plasma column core diminishes as the laser pulse penetrates further
into the sodium vapor.

\begin{figure}[htbp]
\centering
\includegraphics[width=0.85\textwidth]{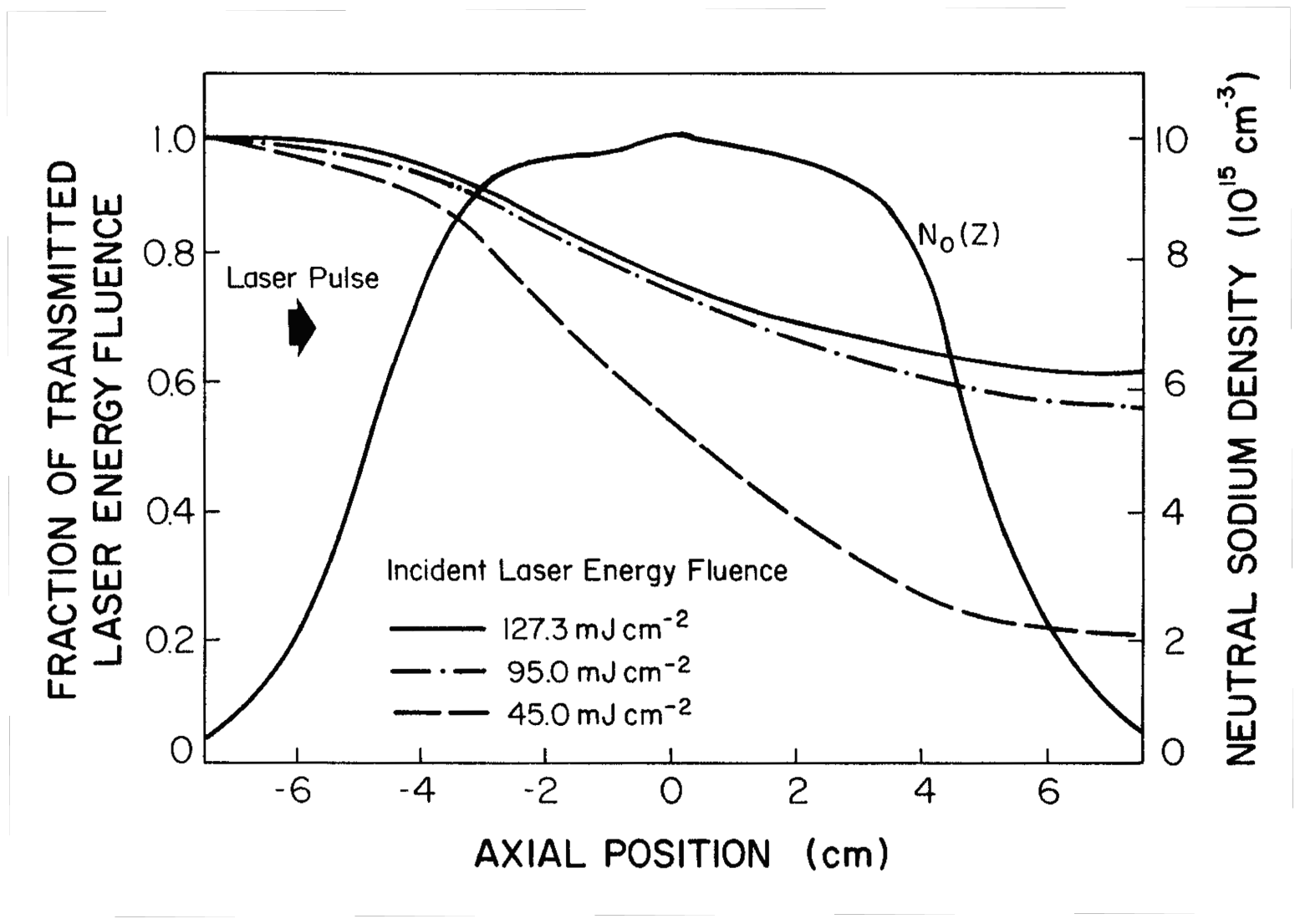}
\caption{Fraction of transmitted laser energy fluence as function of axial
location of four values of incident laser energy fluence. Also shown is the
sodium atom distribution assumed along the path of the laser pulse.}
\label{fig:8}
\end{figure}

This reduction and narrowing of the plasma column is a direct consequence of the
absorption suffered by the laser pulse as it propagates through the sodium vapor
(from $-$ve to $+$ve $z$ values). In figure~\ref{fig:8} we present the axial
variation of the fraction of transmitted laser energy fluence for several
incident values of laser energy fluence for the sodium atom distribution shown in
figure~\ref{fig:7}. It can be seen that the smaller the incident laser energy
fluence the greater its percentage attenuation in propagating through the sodium
vapor. It follows that the radial profile of a laser pulse will tend to steepen as
it propagates since its high intensity core will be proportionately less reduced
than its weaker outer region.

\begin{figure}[htbp]
\centering
\includegraphics[width=0.85\textwidth]{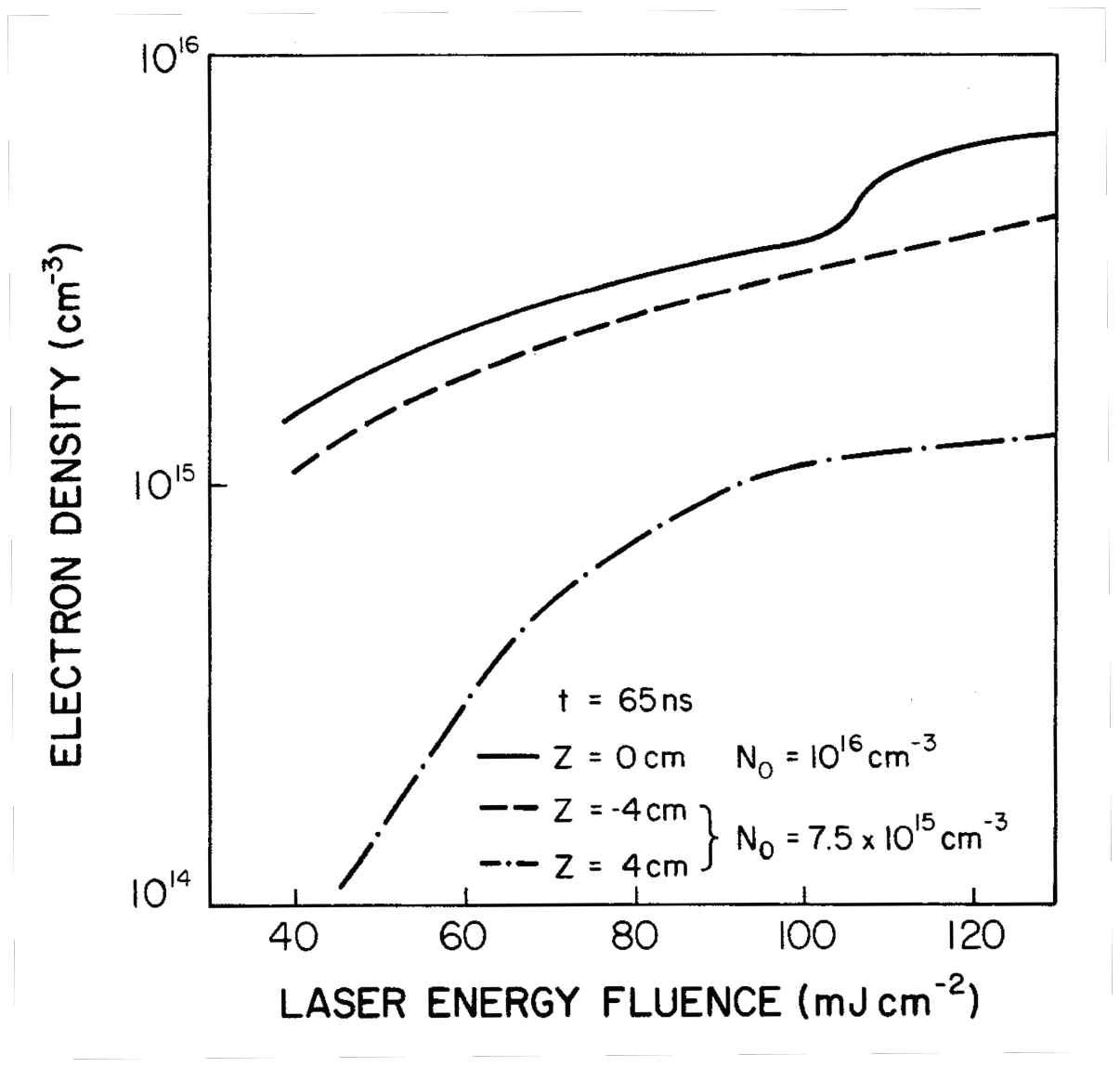}
\caption{Variation of the free electron density (achieved at $Z = -4$, 0 and
4~cm and $t = 65$~ns) with the incident laser energy fluence.}
\label{fig:9}
\end{figure}

This will tend to give rise to an ovoidally shaped region of ionization along the
path of the laser pulse. This is illustrated for the simulation under
consideration in figure~\ref{fig:9}, where the variation of $N_e$ at
$t = 65$~ns and $Z = -4$, 0 and 4~cm are plotted as a function of the incident
laser energy fluence.

\begin{figure}[htbp]
\centering
\includegraphics[width=0.85\textwidth]{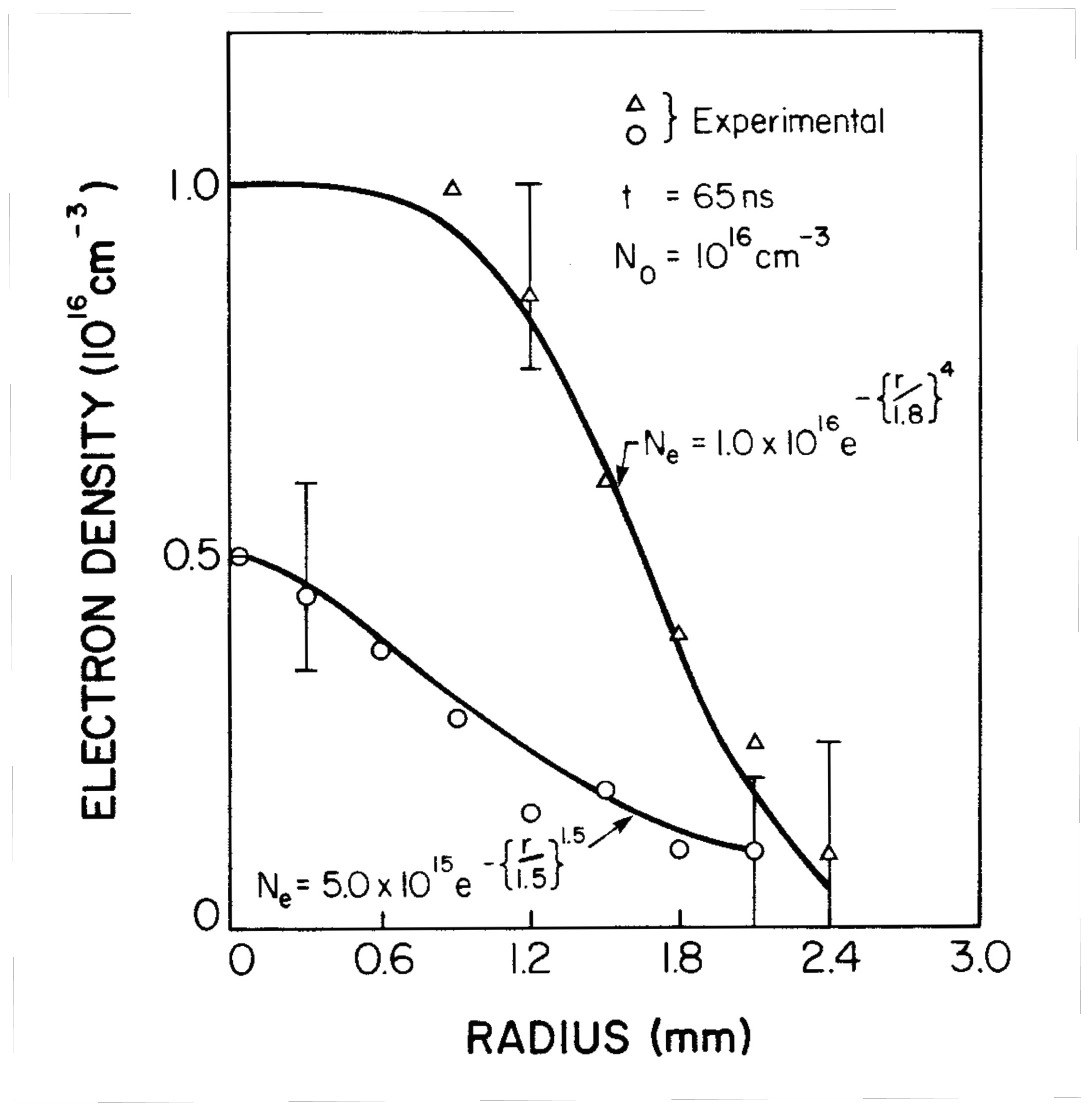}
\caption{Free electron density radial profiles attained at $Z = -2$ and 2~cm for
$N_0 = 10^{16}$~cm$^{-3}$ and $t = 65$~ns, $\varepsilon_{\ell} = 25$~mJ.}
\label{fig:10}
\end{figure}

In figure~\ref{fig:10}, we display the experimentally measured free electron
density radial profiles at $Z = -2$ and 2~cm, 65~ns from the start of the
incident laser pulse. These results were derived by fitting convolved Stark and
instrumental profiles to radially inverted (using the Abel transform)
$4^2D$--$3^2P$ lateral emission spectra\cite{ref37} for the same conditions as
used in the computational work. Also shown in figure~\ref{fig:10} are two
empirical profiles fitted to these data.

\begin{figure}[htbp]
\centering
\includegraphics[width=0.95\textwidth]{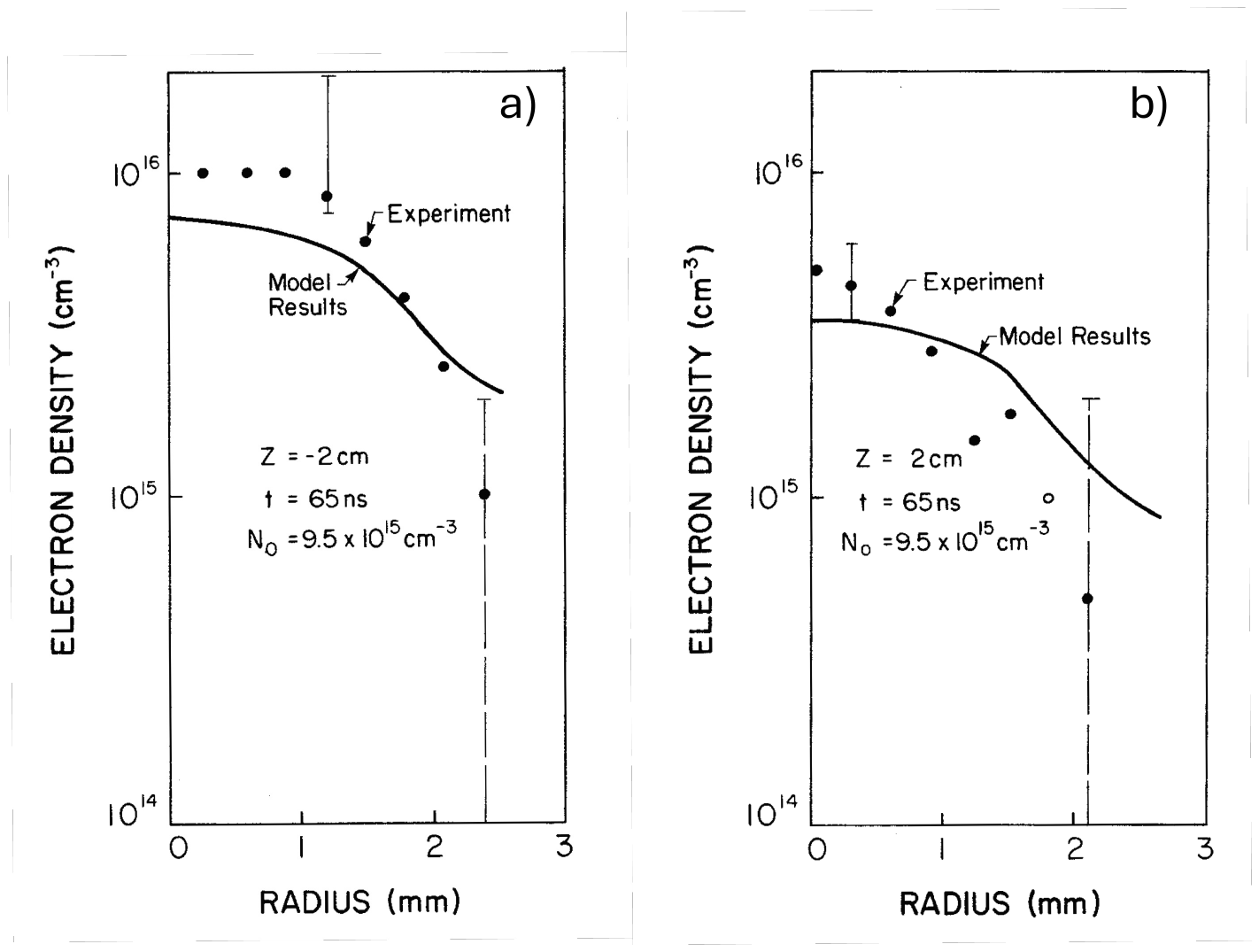}
\caption{Free electron density radial profile at $Z = -2$ and $Z = 2$~cm (b)
where $N_0 = 9.5\times10^{15}$~cm$^{-3}$ and $t = 65$~ns as determined by:
experiment (o) and LIBORS computer code.}
\label{fig:11}
\end{figure}

In figure~\ref{fig:11}(a) we present both the experimental $N_e$-radial profile
(at $Z = -2$~cm and $t = 65$~ns) and the computed profile, while in
figure~\ref{fig:11}(b) we present the corresponding profiles at $Z = 2$~cm. We
should, however, point out that both our sodium atom density measurements and
that of the laser pulse radius have an uncertainty of about
20\%.\cite{ref26} Since the attenuation and temporal and spatial shape distortion
suffered by the laser pulse depends upon these variables it is evident that the
predictions of our computer code would be expected to be of limited accuracy.
Unfortunately the sensitivity of the predictions to uncertainty depend upon the
laser energy fluence so no single figure can be quoted. Nevertheless simulations
suggest that at worst these uncertainties should lead to a factor of two
uncertainty in the code predictions. Reference to figures~\ref{fig:11}(a) and (b)
suggest that the agreement between the experiment and the code results falls
within a factor of two.

\begin{figure}[htbp]
\centering
\includegraphics[width=0.9\textwidth]{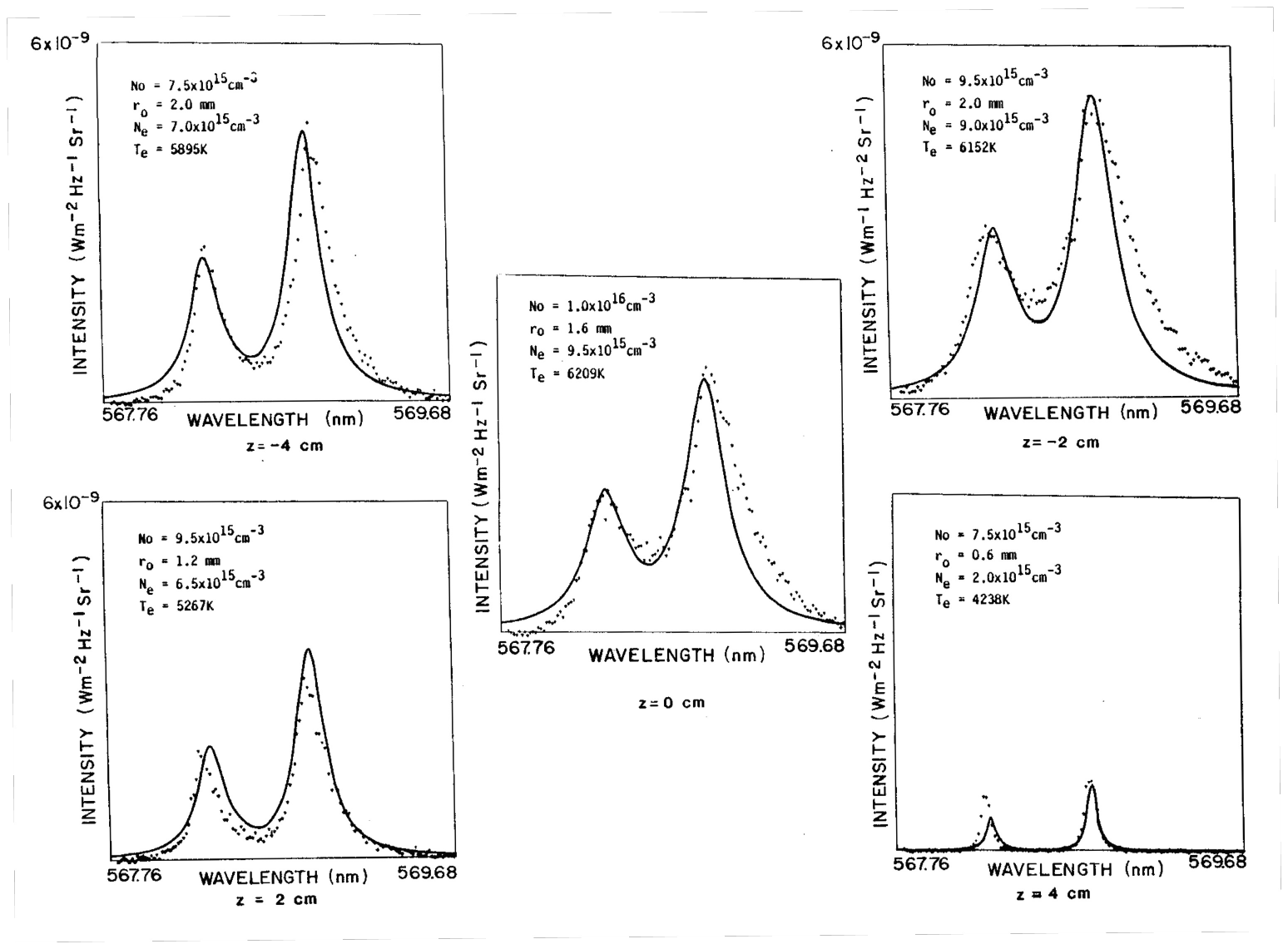}
\caption{Comparison of experimentally measured $4^2D$--$3^2P$ multiplet spectra
at $Z = -4$, $-2$, 0, 2 and 4~cm for a sodium atom distribution with
$N_e^{\mathrm{peak}} = 10^{16}$~cm$^{-3}$ at $t = 65$~ns, where
$E^{\ell} = 25$~mJ, with computer simulation of the same spectra under identical
conditions.}
\label{fig:12}
\end{figure}

In addition to the radial measurements of the free electron density we have also
recorded the $4^2D$--$3^2P$ multiplet spectrum at five ($y = 0$) axial locations
($Z = -4$, $-2$, 0, 2, 4~cm). These spectra can be used to give us a fairly good
idea of the variation of the peak ($r = 0$) electron density along the path of the
laser beam. This is accomplished by matching the spectrum at each $Z$-location to
that evaluated from the one dimensional radiative transfer equation for a uniform
plasma of electron density $N_e$ and radius $r_0$.\cite{ref37} In effect $N_e$ and
$r_0$ are used as fitting parameters. Figure~\ref{fig:12} presents these two sets
of $4^2D$--$3^2P$ multiplet spectra. The computed Stark broadened profiles (full
curves) are seen to approximate the experimental spectra ($\bullet$) in regard
to: the ratio of peak heights, the peak to minimum ratio and the widths of the
profiles.

Confirmation that these spectra, which are integrated along the line of sight for
$y = 0$ (as opposed to the radially inverted spectra used for the radial profile
measurements) can be used to approximately assess the axial ($r = 0$) electron
density is obtained when we compare the predictions at $Z = -2$ and 2~cm from
figure~\ref{fig:12} with the actual radial profiles presented as
figure~\ref{fig:10}.

The axial values of $N_e$ can be seen to decline rapidly for $Z > 0$ and the
asymmetry with regard to equidensity locations is in keeping with our other
observations and with our computational predictions. These results confirm that
attenuation of the laser pulse is a very real effect, even for distances of a few
centimeters, once the atom density is around $10^{16}$~cm$^{-3}$.

\begin{figure}[htbp]
\centering
\includegraphics[width=0.95\textwidth]{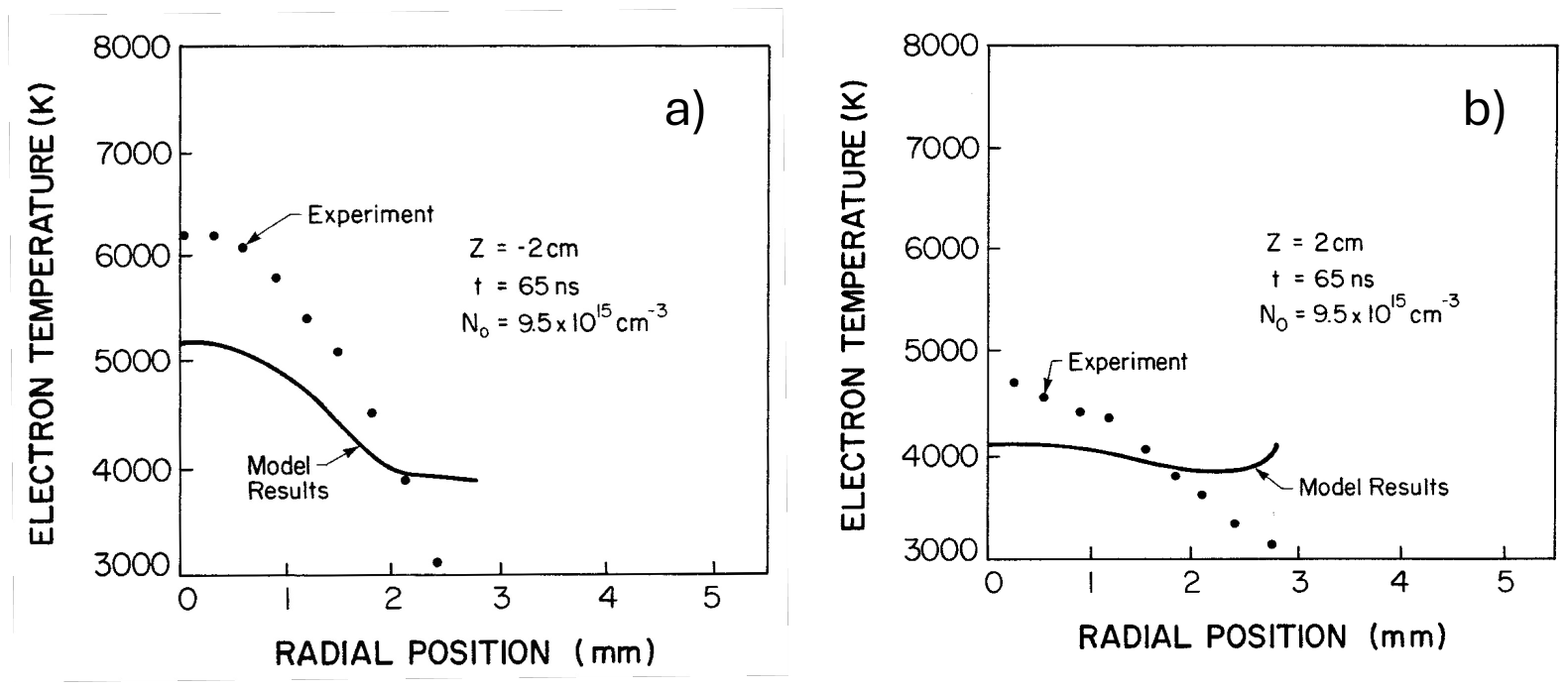}
\caption{Electron temperature radial profile at $Z = -2$~cm (a) and $Z = 2$~cm
(b) where $N_0 = 9.5\times10^{15}$~cm$^{-3}$ and $t = 65$~ns as determined by:
experiment (o) and LIBORS computer code.}
\label{fig:13}
\end{figure}

If we assume that the plasma is in LTE, then the experimentally measured radial
electron density distributions at $Z = -2$ and 2~cm can be used with the Saha
equation and the initial atom density to determine the corresponding radial
profile of the free electron temperature. These are displayed with the
corresponding computed $T_e(r)$ profiles in figure~\ref{fig:13}(a) and (b). In
both cases the experimental data is predicted within a factor of two by the
computational $T_e$-profiles.

\begin{figure}[htbp]
\centering
\includegraphics[width=0.85\textwidth]{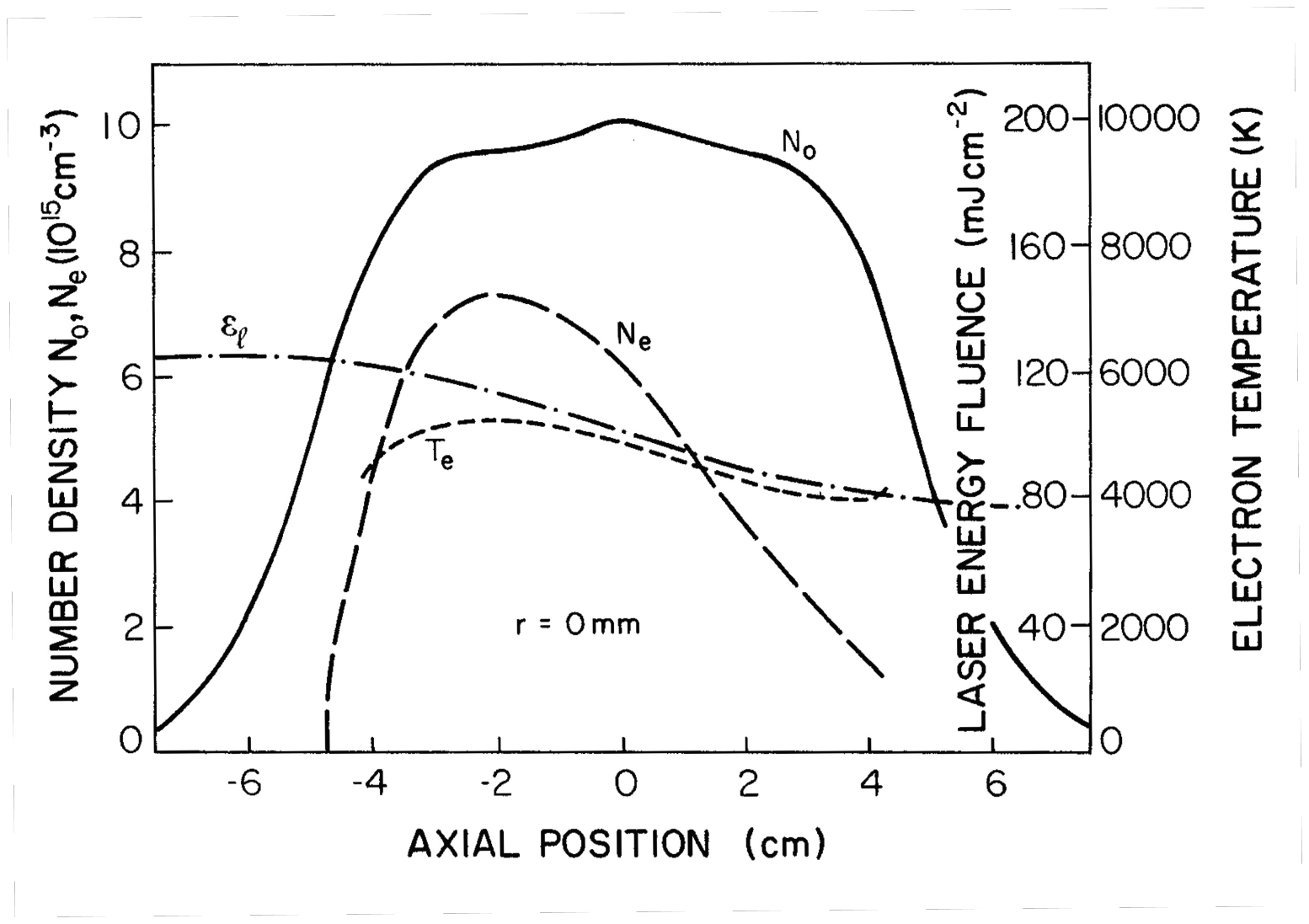}
\caption{Axial variation of the free electron density and temperature as
predicted by LIBORS computer code at $r = 0$ and $t = 65$~ns. Also shown is the
assumed sodium atom density distribution and the decline in the laser energy
fluence with axial location.}
\label{fig:14}
\end{figure}

\begin{figure}[htbp]
\centering
\includegraphics[width=0.85\textwidth]{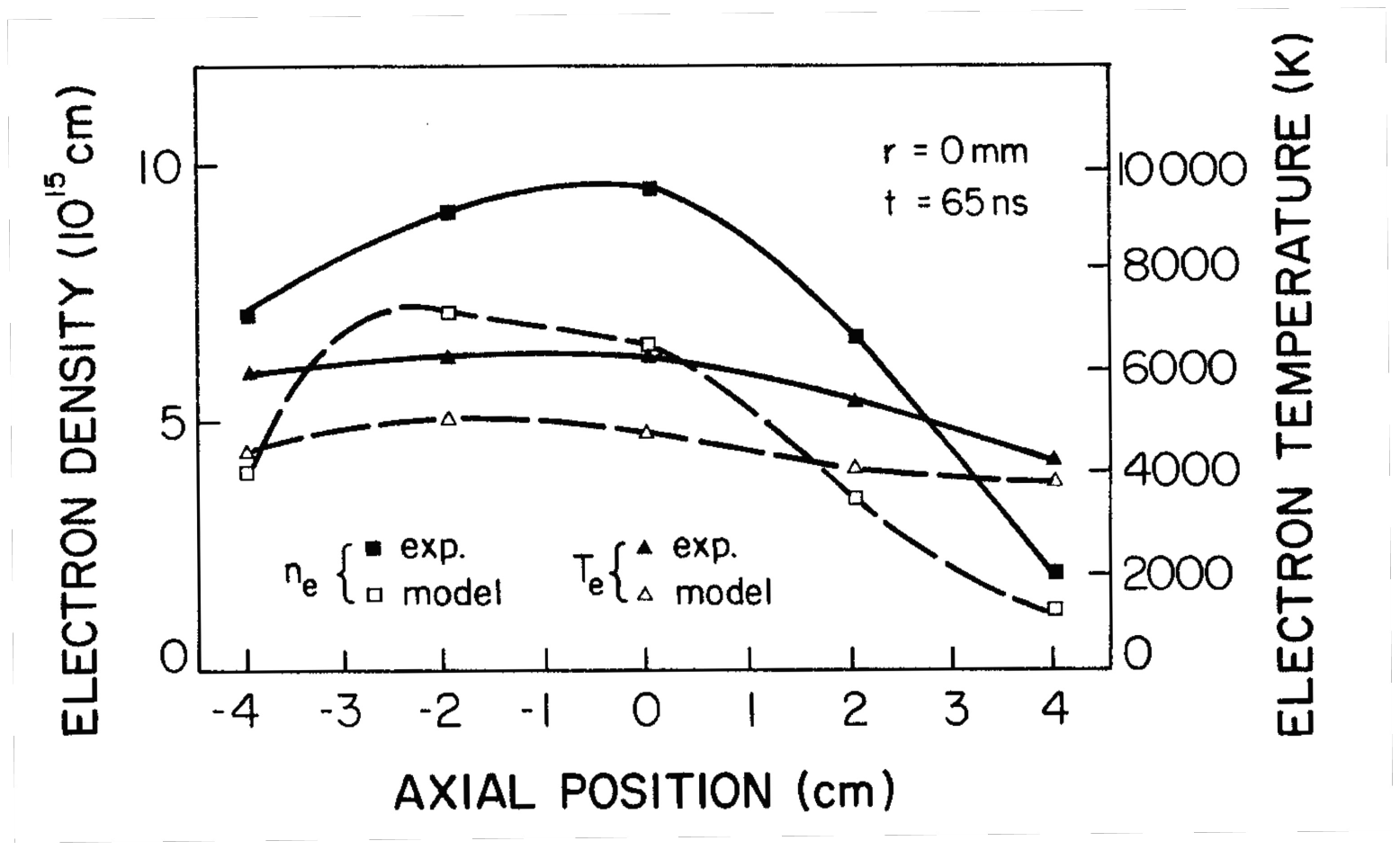}
\caption{Comparison of the axial variation (for $r = 0$) of the free electron
density and temperature determined by experiment and predicted by our LIBORS
computer code.}
\label{fig:15}
\end{figure}

The temperatures indicated in these results are somewhat lower than we had
previously expected from LIBORS but appears to confirm other observations of low
electron temperatures.\cite{ref4,ref5,ref38,ref39} This is particularly so for
$Z = 2$~cm, where the mean temperature is about 4000~K. This low temperature is
also seen to be predicted by our LIBORS code after penetrations of several
centimeters of vapor and represent another manifestation of reduced laser energy
fluence resulting from absorption of the laser pulse. This is clearly seen in
figure~\ref{fig:14} where the computer predicted axial variation of $N_e$ and
$T_e$ (for $r = 0$) are presented with that of $\varepsilon_{\ell}$. Also shown is
the measured axial variation of $N_0$ that was used in the computer code. In
figure~\ref{fig:15} we present a comparison of the computer predicted axial
variation (at $r = 0$) of $N_e$ and $T_e$ with that of the experimentally derived
values based on an assumed radially uniform value of $N_e$ corresponding to the
value estimated at $y = 0$.

\section*{Discussion and Conclusions}

We have developed a computational model of ``laser ionization based on resonance
saturation'' (LIBORS) for sodium vapor that takes into account both the
nonuniform atom distribution and absorption of the laser field. This computer
code has been used to provide the first insight into the three dimensional nature
of this interaction.

We have also undertaken the first measurements of both the radial and axial
profiles of the free electron density in a sodium plasma created by laser
resonance saturation. From these measurements we have deduced the corresponding
radial and axial electron temperature profiles.

We have used our LIBORS computer code to model the sodium plasma formed along the
path of the laser pulse used in our experiments and have compared the code
predicted radial and axial profiles of electron density and temperature with the
results derived from our experiments. In general the agreement is good,
especially if allowance is made for the experimental uncertainties and the limited
accuracy of the many cross-sections used in the code.

We believe that these results demonstrate that our LIBORS computer code is capable
of predicting the three dimensional behaviour of this new mode of laser ionization
with reasonable accuracy (within a factor of two). In particular, both our
computer code and our experiments clearly reveal that absorption of laser energy
due to the strong nature of the interaction leads to a decline in the free
electron density and temperature along the path of the laser pulse. This can, for
modest values of laser energy fluence, lead to quite low values of both $N_e$ and
$T_e$ after the laser pulse has penetrated just a few centimeters of vapor at an
atom density of about $10^{16}$~cm$^{-3}$. This could well explain why some
researchers have failed to observe appreciable ionization\cite{ref40} or have
measured low values of $T_e$.\cite{ref4,ref5,ref38,ref39}

Our results clearly indicate that if LIBORS is to be used to generate fairly
uniform plasma channels, then the laser energy fluence has to be sufficient to
achieve close to full ionization along the path of the atomic vapor. If plasma
channels of several meters length are required,\cite{ref41,ref42} this could
require very large pulses of laser energy and multiphoton interactions might
become significant. Under these circumstances multiphoton ionization might have to
be considered as a viable alternative approach.

In comparing LIBORS with near resonant multiphoton ionization it has to be
recognized that in the former case ionization is achieved through a laser driven
collisional process which invariably leads to populating most of the intermediate
levels~--- all of which lose energy by radiative decay. Near resonant multiphoton
ionization can lead to direct ionization with almost no excitation of intermediate
states for short times. However, it is difficult to achieve full ionization by
this means especially at high densities. Consequently LIBORS may still have some
advantages for creating long, dense, relatively cool plasmas. Clearly more
research is needed to determine the best technique for any given situation.

\section*{Acknowledgements}

This work was supported by the U.S. Air Force Office of Scientific Research
(under grant number AFOSR 85-0020) and the Natural Science and Engineering
Research Council of Canada.


\end{document}